\documentclass{article}

\usepackage{jinstpub} 
\usepackage{inputenc,graphicx}
\usepackage{siunitx}
\usepackage{multirow}
\usepackage{makecell}

\title{Radiation Tolerance of SiGe BiCMOS Monolithic Silicon Pixel Detectors without Internal Gain Layer}

\author[a,1]{M. Milanesio,\note{Corresponding author.}}
\author[a,b]{L. Paolozzi,} 
\author[a]{T. Moretti,}
\author[a]{R. Cardella,}
\author[a]{T. Kugathasan,}
\author[a]{F. Martinelli,}
\author[a,b]{A. Picardi,}
\author[a]{I. Semendyaev,}
\author[a]{S. Zambito,}
\author[d]{K. Nakamura,}
\author[d]{Y. Tabuko,}
\author[d]{M. Togawa,}
\author[c]{M. Elviretti,}
\author[c]{H. Rücker,}
\author[a]{F. Cadoux,}
\author[a,2]{R. Cardarelli\note{Also at INFN Section of Roma Tor Vergata, Via della ricerca scientifica 1, Roma, Italy.},}
\author[a]{S. Débieux,}
\author[a]{Y. Favre,}
\author[a]{C. A. Fenoglio,}
\author[a]{D. Ferrere,}
\author[a]{S. Gonzalez-Sevilla,}
\author[a]{L. Iodice,}
\author[a,b]{R. Kotitsa,}
\author[a]{C. Magliocca,}
\author[a,b]{M. Nessi,}
\author[a]{A. Pizarro-Medina,}
\author[a]{J. Sabater Iglesias,}
\author[a]{J. Saidi,}
\author[a]{M. Vicente Barreto Pinto,}
\author[a,1]{and G. Iacobucci}
\affiliation[a]{D\'epartement de Physique Nucl\'eaire et Corpusculaire (DPNC),
University of Geneva, 24 Quai Ernest-Ansermet, CH-1211 Geneva 4, Switzerland}

\affiliation[b]{CERN, CH-1211 Geneva 23, Switzerland}

\affiliation[c]{IHP — Leibniz-Institut für innovative Mikroelektronik, Im Technologiepark 25, Frankfurt (Oder), Germany}

\affiliation[d]{High Energy Accelerator Research Organization, Oho 1-1, Tsukuba-shi, Ibaraki-ken, Japan}

\emailAdd{matteo.milanesio@unige.ch, giuseppe.iacobucci@unige.ch}

\abstract{
A  monolithic silicon pixel prototype  produced for the MONOLITH ERC Advanced project was irradiated with 70 MeV protons up to a fluence of $1\times 10^{16}$ 1 MeV n$_{\text{eq}}$/cm$^{2}$. The ASIC contains a matrix of hexagonal pixels with 100 $\mu$m pitch, readout by low-noise and very fast SiGe HBT frontend electronics. Wafers with 50 $\mu$m thick epilayer with a resistivity of 350 $\Omega$cm  were used to produce a fully depleted sensor. 
Laboratory tests conducted with a $^{90}$Sr source show that the detector works 
satisfactorily after irradiation.
The signal-to-noise ratio is not seen to change up to fluence of  $6 \times 10^{14}$ n$_{\text{eq}}$/cm$^{2}$.
The signal time jitter 
was estimated as the ratio between the voltage noise and the signal slope at threshold.
At -35$^\circ$C, sensor bias voltage of 200 V and frontend power consumption of  0.9 W/cm$^2$,
the time jitter of the most-probable signal amplitude was estimated to be
$\sigma_{t}^{^{90}\!{\text{Sr}}}$  = 21 ps
for proton fluence up to $6 \times 10^{14}$ n$_{\text{eq}}$/cm$^{2}$
and 57 ps at $1 \times 10^{16}$ n$_{\text{eq}}$/cm$^{2}$.
Increasing the sensor bias to 250 V and the analog voltage of the preamplifier from 1.8 to 2.0 V provides a time jitter of 40 ps at $1 \times 10^{16}$ n$_{\text{eq}}$/cm$^{2}$.

}

\begin{document}
\maketitle
\section{Introduction}
\label{sec:intro}

Recent researches in the framework of the MONOLITH Horizon2020 ERC Advanced project~\cite{monolith} demonstrated that SiGe HBT electronics can be used to produce low noise, low power and very fast frontend that could be integrated in a monolithic ASIC to obtain a fully efficient detector with excellent time resolution.

Several ASICs were produced using the SG13G2 process~\cite{SG13G2} by IHP microelectronic and characterized at the SPS testbeam  facility at CERN with 120 GeV/c pions.
In a first ASIC version with an internal gain layer~\cite{PicoADpatent,picoad_gain}, time resolutions of 17 ps were measured~\cite{PicoAD_TB} 
with a dependence on the hit position varying from 13 ps at the center of the pixel and 25 ps at the inter-pixel region.

More recently, a second prototype was produced with improved electronics. A version without internal gain layer~\cite{Zambito_2023} provided 20 ps time resolution, with  little dependence on the  position of the hit in the pixel area.
To study the radiation tolerance of SiGe HBT\footnote{Previous studies on SiGe HBT radiation tolerance can be found in~\cite{Ullan, Cressler} and references therein.}
 in view of applications to future colliders, eight samples of this second monolithic prototype
were irradiated with 70 MeV protons up to $1 \times 10^{16}$ n$_{\text{eq}}$/cm$^{2}$. 
In this paper, we present the results of laboratory measurements performed with radioactive sources to assess the timing performance of the SiGe HBT amplifier implemented in the chip prototype described in~\cite{Zambito_2023}. 
The measurements focus on the characterization of the single-ended output of four analog pixels,  consisting of a fast charge amplifier in SiGe HBT and a two-stages analog driver that allows for direct measurement of the analog pulse using a fast oscilloscope.

\section{Proton irradiation of the SiGe HBT ASICs}
\label{sec:results}

Eight prototype chips of the same type that have been studied in~\cite{Zambito_2023} were wire-bonded on readout boards and characterized with a $^{55}$Fe radioactive source in the clean rooms of the University of Geneva, in terms of gain and  Equivalent Noise Charge (ENC). 
The eight boards were then shipped to the Cyclotron and Radioisotope Center (CYRIC) proton irradiation facility at Tohoku University in Japan, which has an azimuthal-varying field cyclotron and a beamline for the radiation damage test of semiconductors \cite{Nakamura_2015}. The beamline supplies a high-intensity proton beam with a kinetic energy of 70 MeV and a beam current of 1.5 $\mu$A. The eight chips were irradiated with proton fluences varying from $2 \times 10^{13}$ n$_{\text{eq}}$/cm$^{2}$ to $1 \times 10^{16}$ n$_{\text{eq}}$/cm$^{2}$. The electronics on the chips was powered during irradiation. 
A 15-mm-thick aluminum mask was positioned in front of the boards to shield the active components surrounding the chips. 

\begin{figure}[!htb]
    \centering
    \includegraphics[width=0.65\textwidth]{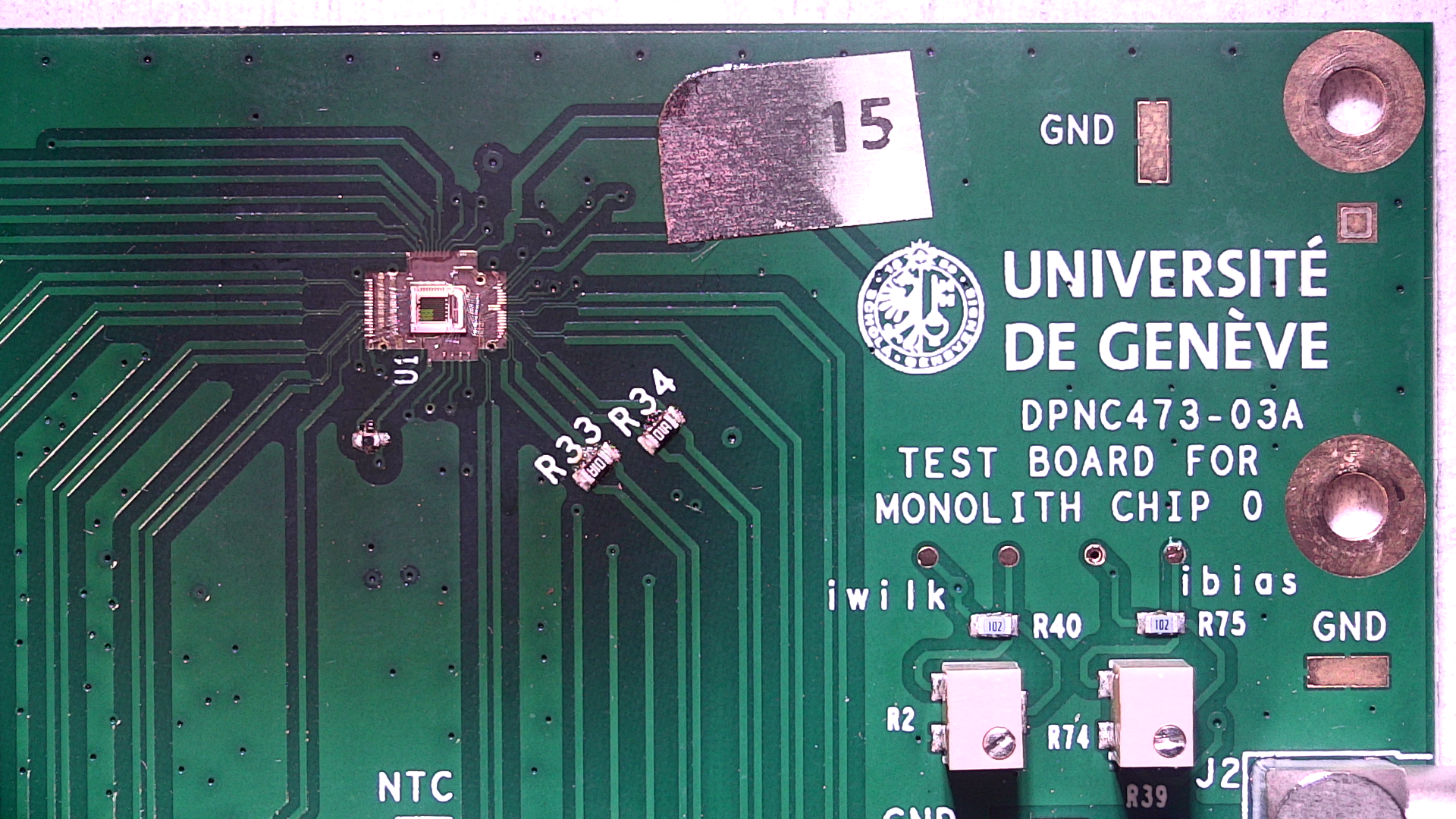}
    \hfill
    \caption{Photograph of a board irradiated with a proton fluence of $1 \times 10^{16}$ n$_{\text{eq}}$/cm$^{2}$. The area around the chip was   burned by radiation and some electronics components of the board were severely damaged.}
    \label{fig:M15_irrad}
\end{figure}
Figure \ref{fig:M15_irrad} shows the readout board hosting one of the two chips irradiated at the maximal fluence of $1\times 10^{16}$ n$_{\text{eq}}$/cm$^{2}$. 
Since not all the protons were  stopped by the shielding,  the radiation damaged some of the Low-Dropout Regulators (LDOs) used to generate low-voltage biases to the chip. A clear correlation was found between the number of damaged LDOs, their position on the board, and the fluence of irradiation. To bypass damaged LDOs, copper wires were soldered to the chip to supply directly the correct low voltage via a power supply.

After irradiation, the chips were tested again in the clean rooms of the University of Geneva. 
One of the two chips irradiated at $3 \times 10^{15}$ n$_{\text{eq}}$/cm$^{2}$ presented a short circuit on the digital power supply, and it was no longer usable after irradiation. This might be related to radiation damage on the chip itself or to other causes related to transport, storage, or manipulation of the chip. One likely cause could be the failure of the LDOs on the board during irradiation, which set a voltage of 3.0 V on the chip, well above the 1.2 V specified for the LVMOS in SG13G2. 
This chip was excluded from the following analysis.

As a consequence, a total of seven irradiated chips were considered for this study. 
In addition, three unirradiated chips were characterized (one of them is the same studied in~\cite{Zambito_2023}) and were used to have a reference for the behavior of the ASIC in the absence of radiation. 
Table \ref{tab:fluences} lists the five proton fluences considered, the number of boards, and the number of characterized analog pixels for each irradiation point.
\begin{table}[!htb]
\centering
\renewcommand{\arraystretch}{1.1}
\begin{tabular}{|c|c|c|c|}
\cline{1-4}
\cline{1-4}
\ Fluence  & Boards &  Boards & Analog pixels \\
\ [1 MeV n$_{\text{eq}}$/cm$^{2}$] & ~prepared~  &  ~~utilised~~  & characterised  \\
\cline{1-4}
0 & 3 & 3 & 12 \\
\cline{1-4}
$2 \times 10^{13}$ & 1 & 1 & 4 \\
$9 \times 10^{13}$ & 1 & 1 & 4 \\
$6 \times 10^{14}$ & 2 & 2 & 8 \\
$3 \times 10^{15}$ & 2 & 1 & 4 \\
$1 \times 10^{16}$ & 2 & 2 & 8 \\
\cline{1-4}
\end{tabular}
\caption{Summary of the boards prepared for this study and of the proton fluence at which they were exposed. The last two columns report the number of boards and analog pixels that were used.}
\label{tab:fluences} 
\end{table} 

During transportation,  storage in the clean rooms and measurement with the radioactive sources, the temperature of the chips was kept to -$35\ ^{\circ}$C to avoid any unwanted annealing.
\subsection{Consequence of proton irradiation on the detector working point}
\label{subsec:consequenceirrad}

The working point of the frontend is defined by two currents that can be controlled via software: the preamplifier current $I_{\it{preamp}}$ and the feedback current $I_{\it{fbk}}$. 

\begin{figure}[!htb]
    \centering
    \includegraphics[width=0.65\textwidth]{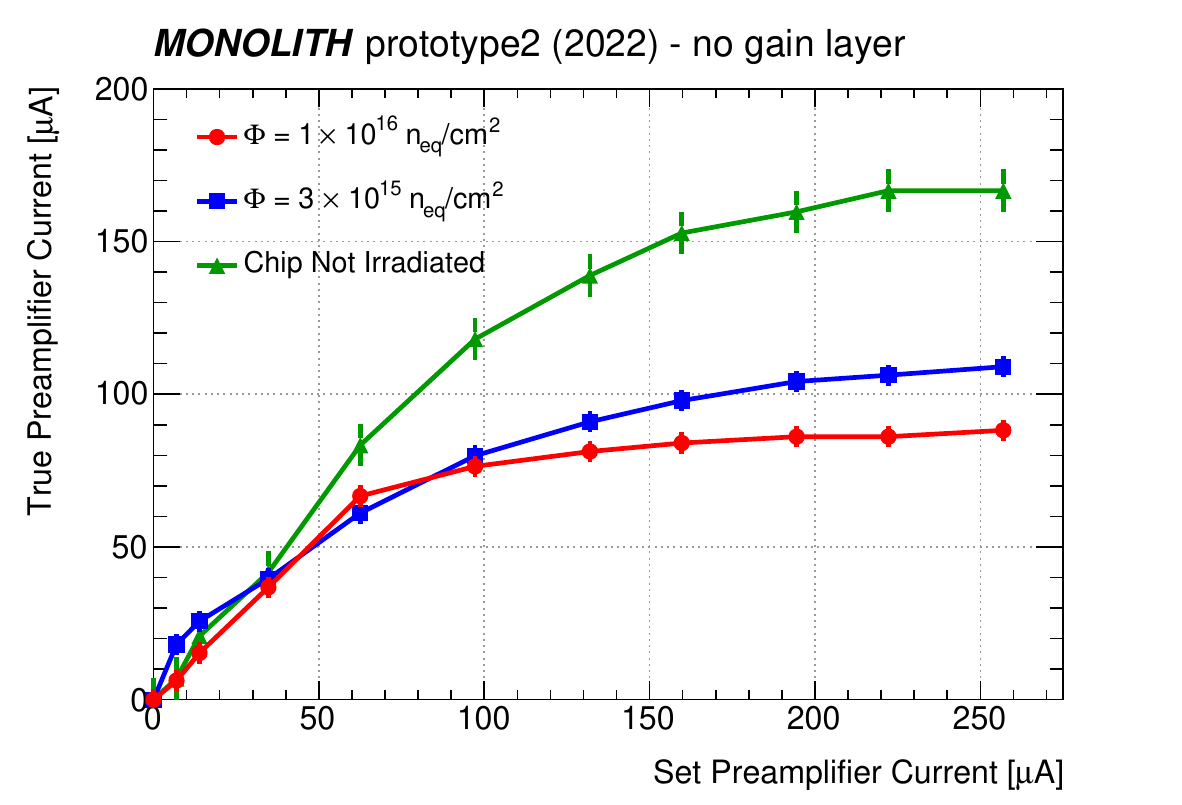}
    \caption{True preamplifier current vs. set preamplifier current for three different proton fluences. The level of saturation of the true preamplifier current shows that the maximum current available to polarize the transistor diminishes with increasing fluence. A set current of 50 $\mu$A, that corresponds to $P_{\it density}$ = 0.9\ W/cm$^{2}$,  was chosen for all the measurements of this study. }
    \label{fig:true-vs-set-current}
\end{figure}

The preamplifier current $I_{\it{preamp}}$ is responsible for supplying the bias current to the collector of the HBT and setting the gain-bandwidth product of the HBT. It also determines the power consumption of the preamplifier, whose power density
can be calculated as:
\begin{equation}
    P_{\it density} = I_{\it preamp}\cdot V_{\it CCA}\cdot N_{\it pixel},
\end{equation}
where $V_{\it CCA}$ is the analog voltage given to the preamplifier (in the following, if not indicated, it is 1.8 V), and $N_{\it pixel}$ is the number of pixels per square cm.
It was observed that radiation damage limits the range in which it is possible to set the preamplifier current. 
The change in characteristics of the bias generation circuit can be easily verified by checking that the measured current absorption of the analog power supply (\textit{true preamplifier current}) be equal to the current provided to the preamplifier (\textit{set preamplifier current}). 
As shown in Figure~\ref{fig:true-vs-set-current}, it was found that the curve of true current vs. set current saturates rapidly at high proton fluence.
Consequently, it was decided to operate all chips at $I_{\it preamp}$ = 50 $\mu$A to compare at the same power consumption chips subject to different proton fluence.
This current corresponds to a power density $P_{\it density}$ = 0.9\ W/cm$^{2}$ at which all chips could be operated, in contrast to the maximum value $P_{\it density}$ = 2.7\ W/cm$^{2}$ that can be achieved before irradiation~\cite{Zambito_2023}.


The feedback current $I_{\it{fbk}}$ is, on the other hand, responsible for supplying the bias current to the base of the HBT and for setting the working point in the gain-bandwidth curve determined by the value of $I_{\it preamp}$. Thus, it changes the negative feedback of the preamplifier, hence regulating its gain and speed. 
Since radiation damage causes an increase of the base leakage current of the HBT,
the working point of the frontend needs to be adjusted.
The best performance is achieved by increasing the current supplied to the HBT base up to the point in which the bias generation circuit is not able to provide the extra feedback current anymore\footnote{The limit to the maximum deliverable current from the biasing circuit is related to the circuit design, and not to a technology limitation.}, as observed for $I_{\it{preamp}}$. To choose a  working point suitable for all the chips, scans as a function of the $I_{\it{fbk}}$ were conducted prior to data taking with the radioactive source. 
The value $I_{\it fbk}$ = 2.0 $\mu$A was chosen, which allowed satisfactory operation of all the irradiated chips. This feedback current value was used to operate also the three unirradiated chips, in contrast to the value $I_{\it fbk}$ = 0.1 $\mu$A used for one of the unirradiated chips in~\cite{Zambito_2023}.

\subsection{Consequence of proton irradiation on the signals and the noise}
\label{subsec:radioactiveSourceChoice}

$^{90}$Sr is an almost pure $\beta^{-}$ emitter, with two main decays with energies of $0.55$ MeV and $2.28$ MeV. Most of this source's electrons behave similarly to minimum ionizing particles in a silicon sensor with $50\ \mu m$ depletion, such as this MONOLITH prototype.

\begin{figure}[!htb]
    \centering
    \begin{minipage}[b]{0.33\textwidth}
        \includegraphics[width=\textwidth]{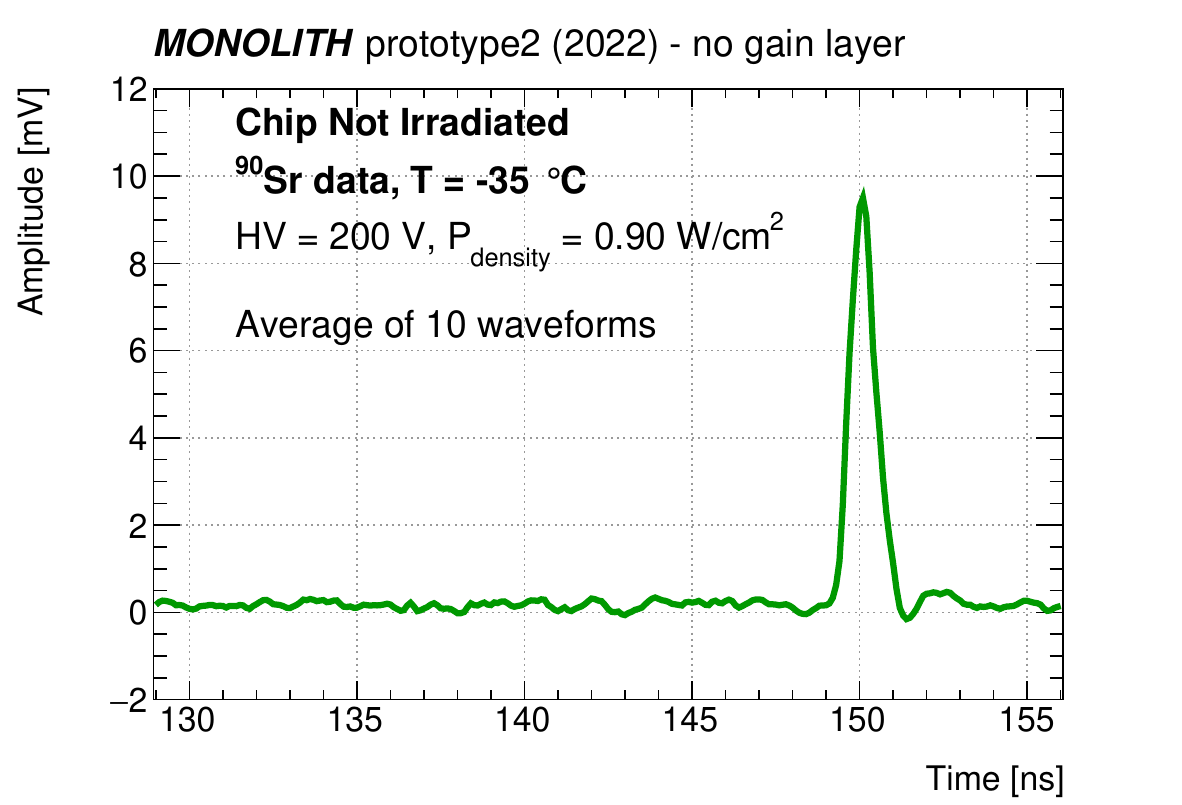}
    \end{minipage}
    \hfill
    \begin{minipage}[b]{0.33\textwidth}
        \includegraphics[width=\textwidth]{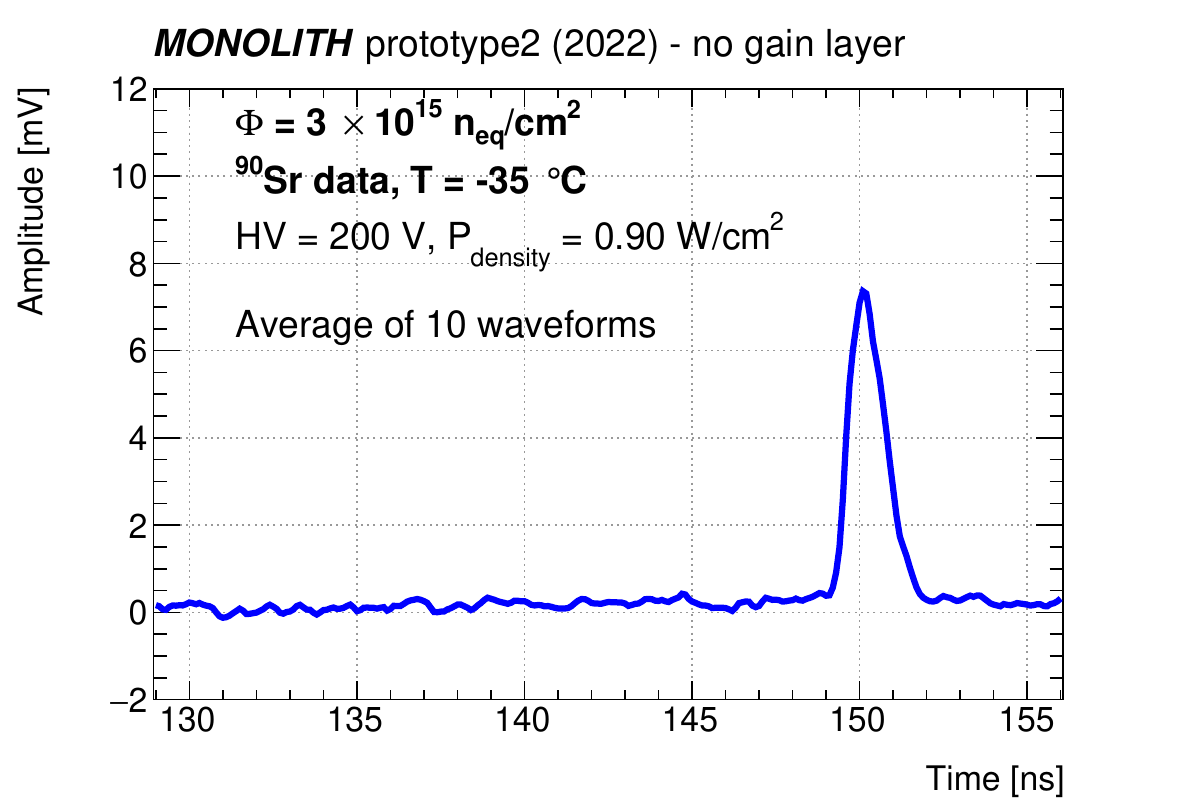}
    \end{minipage}
    \hfill
    \begin{minipage}[b]{0.33\textwidth}
        \includegraphics[width=\textwidth]{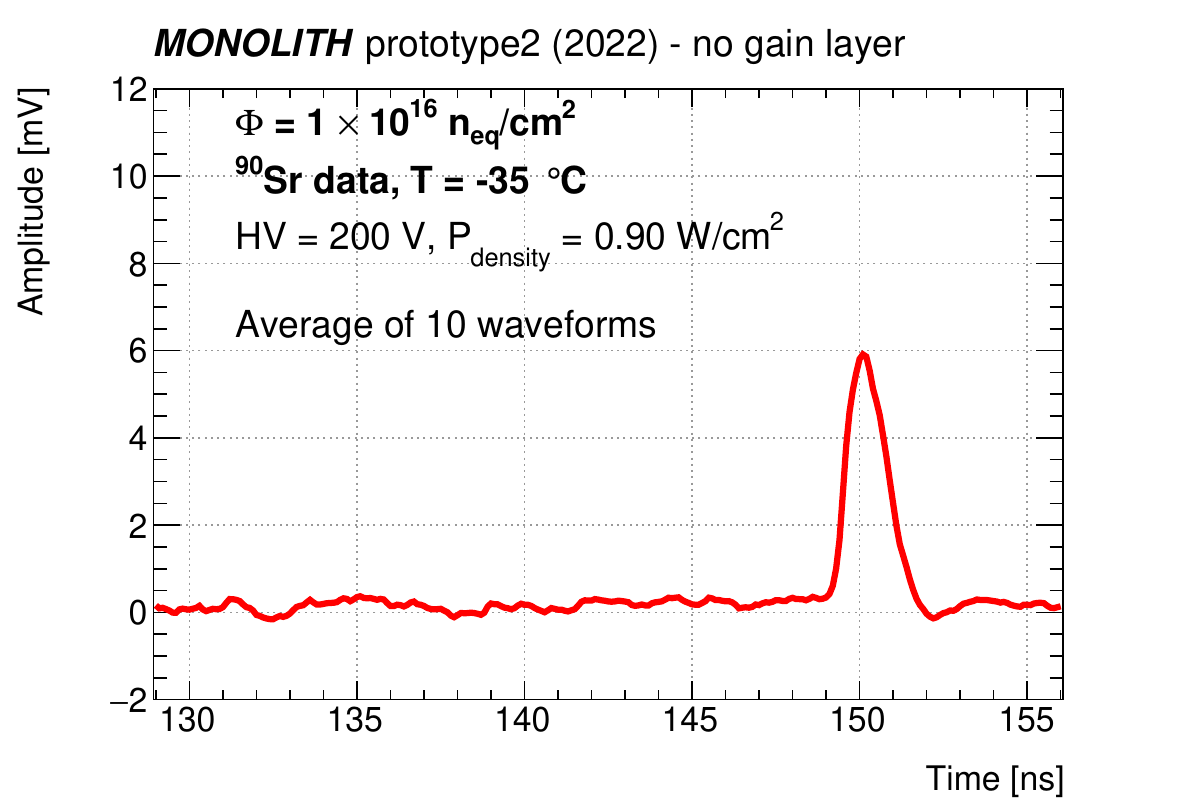}
    \end{minipage}
    \begin{minipage}[b]{0.33\textwidth}
        \includegraphics[width=\textwidth]{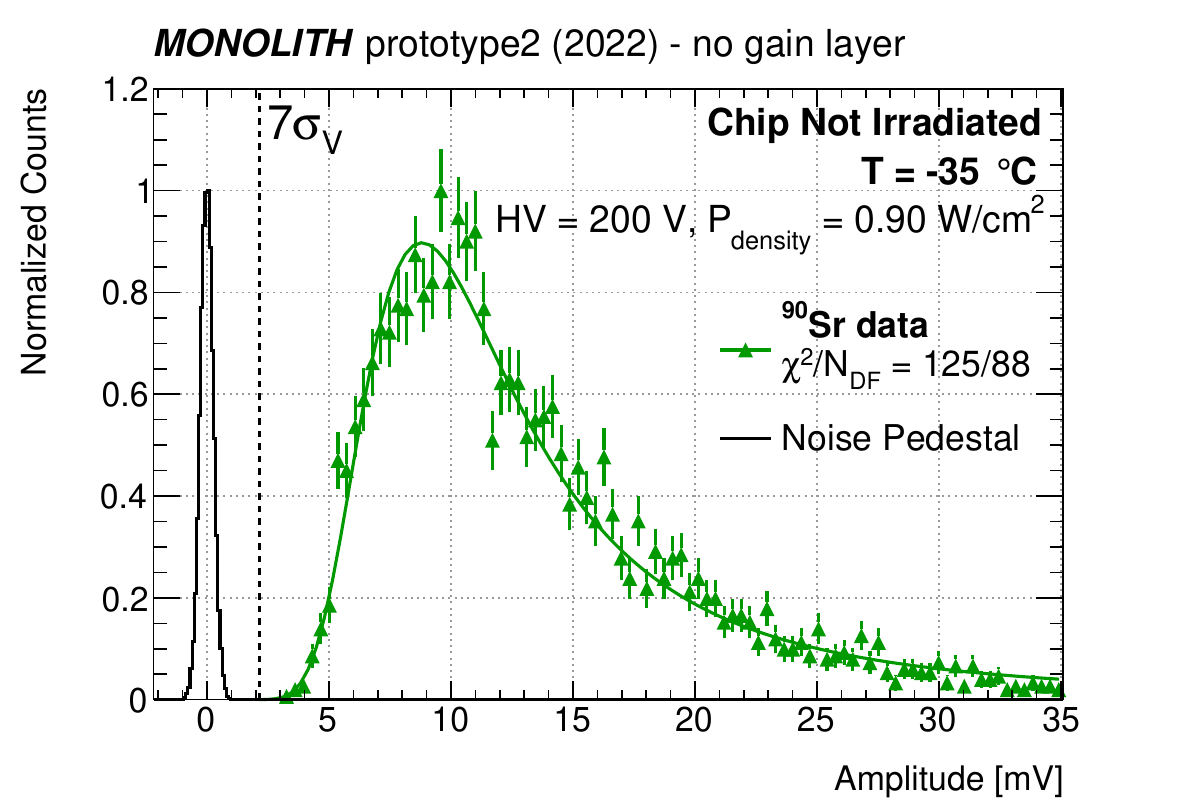}
    \end{minipage}
    \hfill
    \begin{minipage}[b]{0.33\textwidth}
        \includegraphics[width=\textwidth]{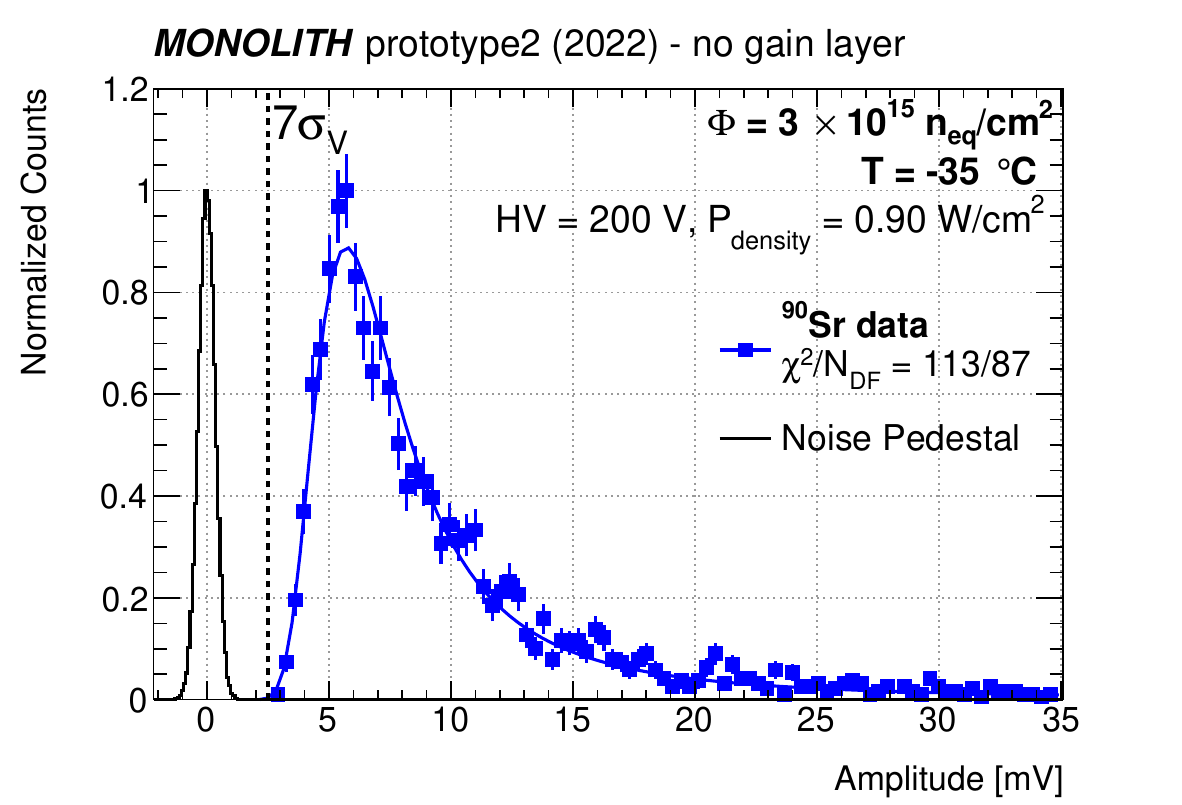}
    \end{minipage}
    \hfill
    \begin{minipage}[b]{0.33\textwidth}
        \includegraphics[width=\textwidth]{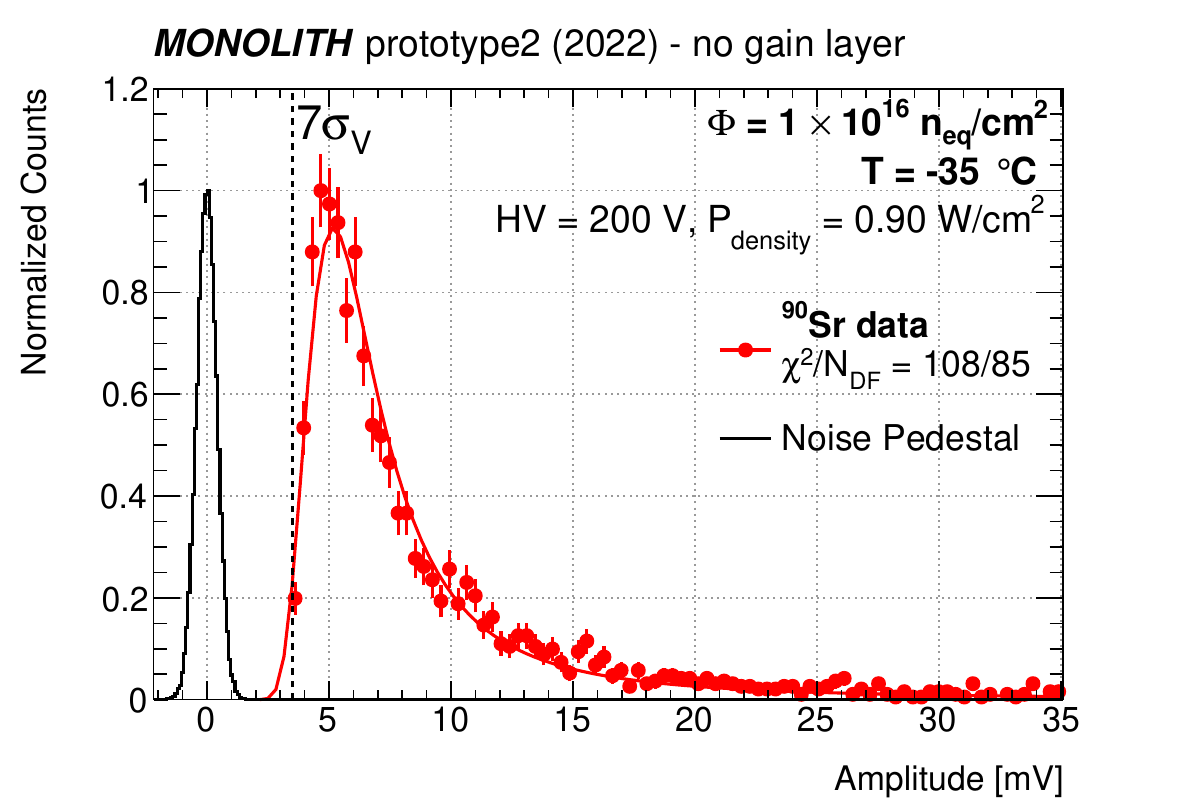}
    \end{minipage}
    \caption{
    Waveforms (top) and amplitude distributions (bottom) acquired using a $^{90}$Sr radioactive source at a preamplifier power consumption of 0.9 W/cm$^2$, for
    an unirradiated (left), irradiated at $3 \times 10^{15}$ n$_{\text{eq}}$/cm$^{2}$ (center) and  $1 \times 10^{16}$ n$_{\text{eq}}$/cm$^{2}$ (right) chip.
%
 %
The waveforms are the average of 10 signals, selected to have an amplitude within $\pm$ 5 mV from the mode of the Landau distributions. 
The colored lines superimposed to the amplitude distributions are the results of fits using a
Landau functional form, which were used to obtain the values of the mode of the distributions; the bottom panels also display the noise pedestals (black histograms) as well as the amplitude that corresponds to the 7 $\sigma_V$ threshold cut (dashed vertical lines).
    }
    \label{fig:waveforms}
\end{figure}

The three top panels in Figure~\ref{fig:waveforms} show the average of 10 signals acquired by a 1 GHz bandwidth oscilloscope from one of the single-ended analog outputs in the case of a chip not  irradiated, and one of the chips irradiated with fluences of $3 \times 10^{15}$  and $1 \times 10^{16}$ n$_{\text{eq}}$/cm$^{2}$. 
The 10 signals were selected with amplitude within $\pm$ 5 mV from the mode of the Landau distributions.
The average signal amplitude from the chip not irradiated is 9 mV; it decreases to approximately 6 mV at $1 \times 10^{16}$ n$_{\text{eq}}$/cm$^{2}$.

The signal-amplitude distributions for the same three pixels are shown in the bottom panels of Figure~\ref{fig:waveforms}, where  the results of the fits using a Landau functional form are superimposed. The figures display also the noise pedestals which were used to determine the voltage noise $\sigma_V$ that was used to set the 7 $\sigma_V$ threshold used for the analysis.
The dashed vertical lines display the position of the threshold in each case, indicating that only in the case of the chip irradiated to $1 \times 10^{16}$ n$_{\text{eq}}$/cm$^{2}$ the voltage threshold cut into the amplitude distribution, creating an inefficiency at the level of a few percent.

\begin{figure}[!htb]
    \centering
    \begin{minipage}[b]{0.49\textwidth}
    \includegraphics[width=\textwidth]{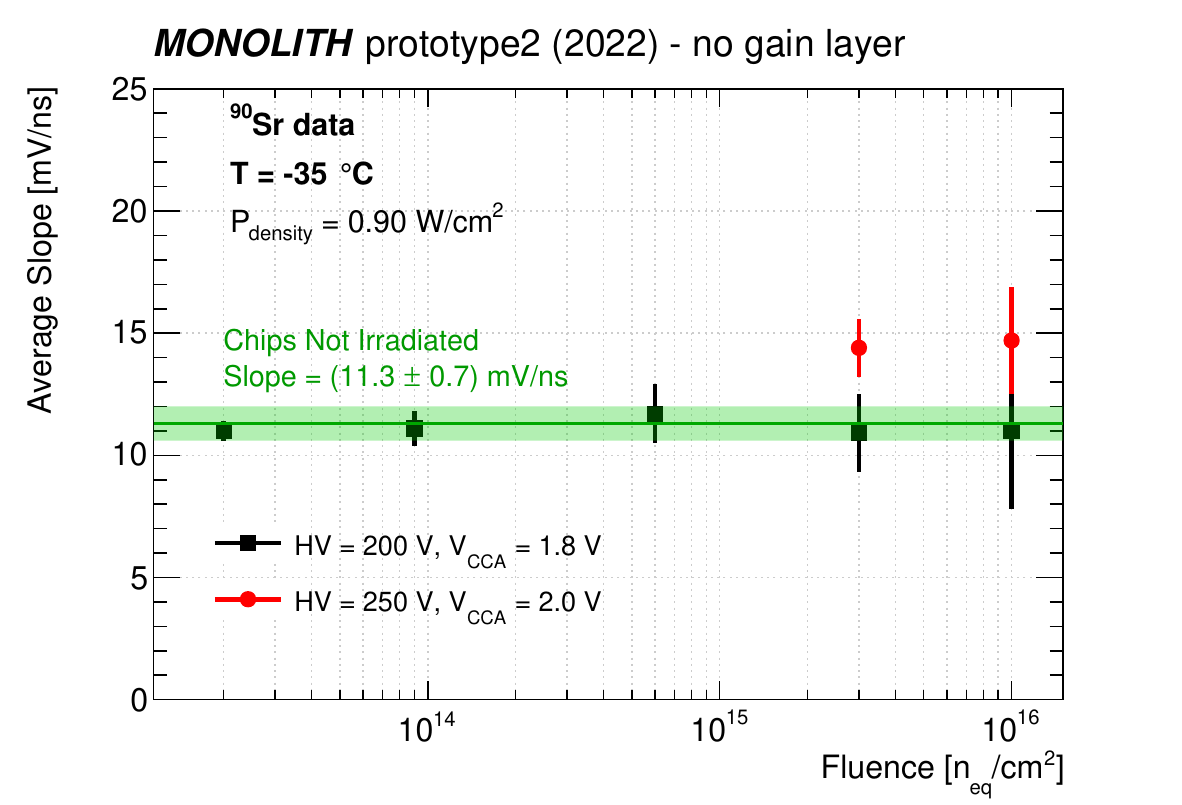}
    \end{minipage}
    \centering
    \hfill
    \begin{minipage}[b]{0.49\textwidth}
        \includegraphics[width=\textwidth]{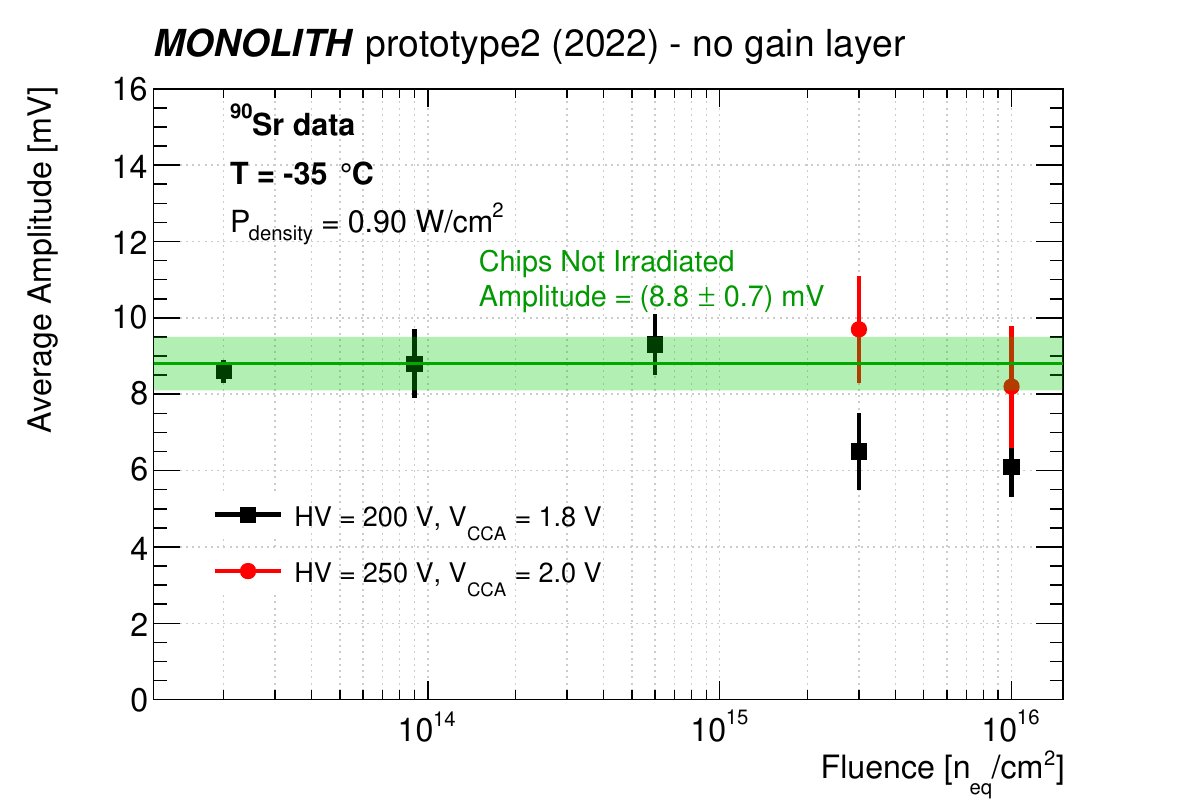}
    \end{minipage}
    \centering
    \hfill
    \begin{minipage}[b]{0.49\textwidth}
        \includegraphics[width=\textwidth]{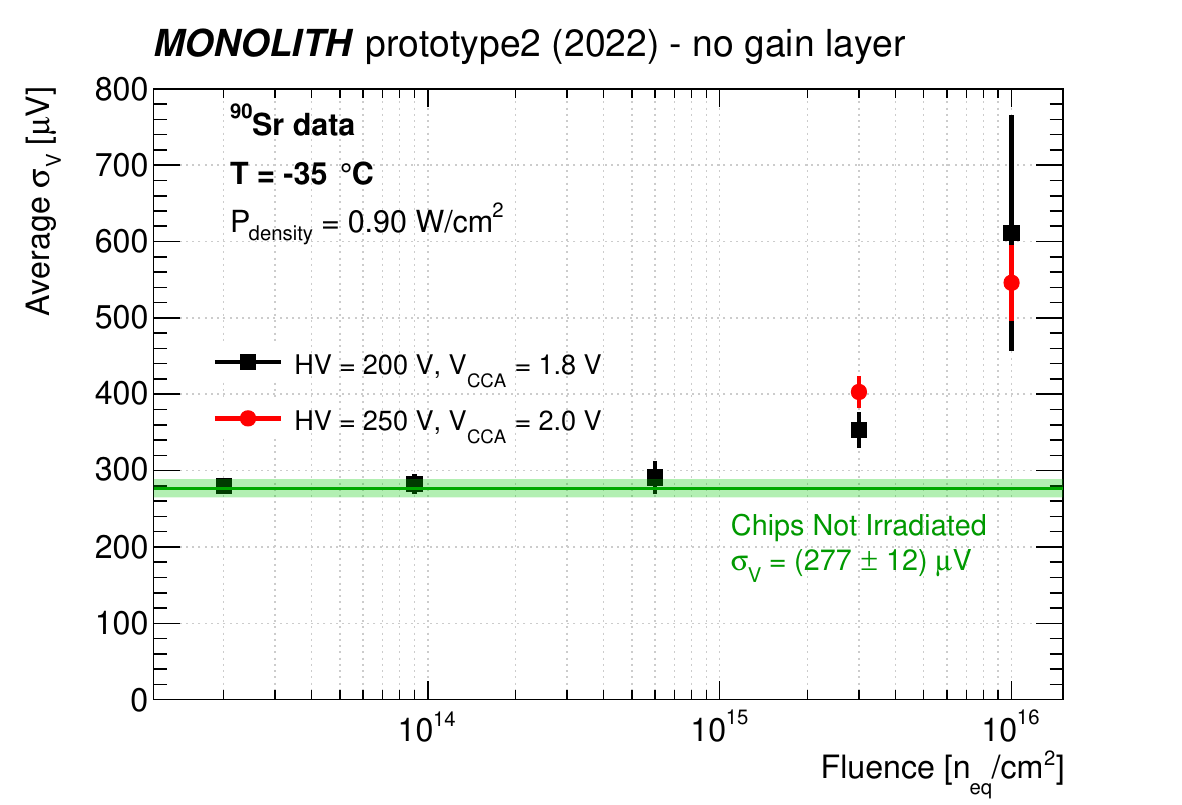}
    \end{minipage}
    \centering
    \hfill
    \begin{minipage}[b]{0.49\textwidth}
        \includegraphics[width=\textwidth]{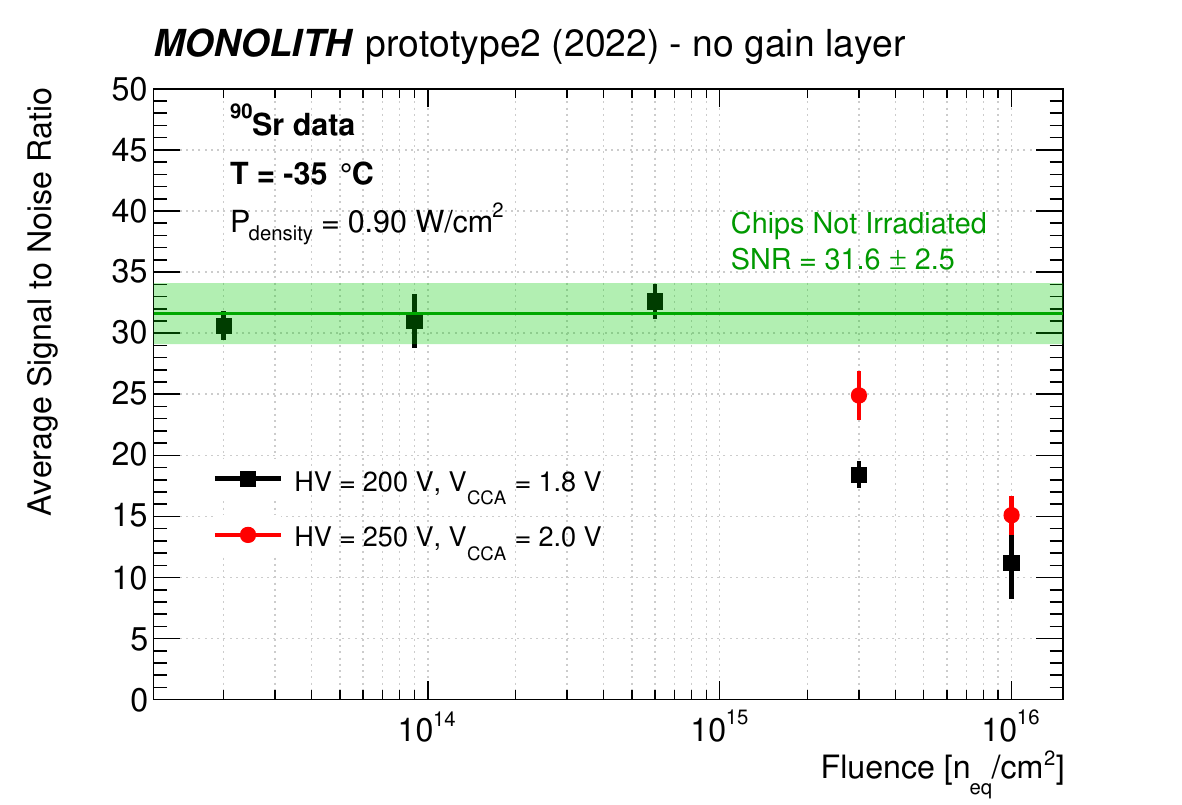}
    \end{minipage}
    \caption{Average signal and noise characteristics measured with the $^{90}$Sr radioactive source as a function of the fluence:     
    the average signal slope at the 7 $\sigma_V$ threshold value (top-left), the average mode of the signal amplitude (top-right), the average $\sigma_{V}$ (bottom-left), and the average signal-to-noise ratio (bottom-right). Data were collected at -35$^\circ$C and $P_{\it density}$ = 0.9 W/cm$^2$, and with two different working points: HV = 200 V and $V_{\it CCA}$ = 1.8 V (black squares), or HV = 250 V and $V_{\it CCA}$ = 2.0 V (red dots). The error bars displayed represent the standard deviation of the average values. 
    The green lines represent the average value measured with the 12 analog pixels of the three unirradiated chips operated in the same conditions as the black squares; the green bands represent the standard deviation of the average values.}
    \label{fig:fluencePlots}
\end{figure}

Figure \ref{fig:fluencePlots} presents the main characteristics of the signals produced by the electrons emitted by the $^{90}$Sr radioactive source, namely the signal amplitude (which depends on the amplifier gain), the voltage noise $\sigma_V$, the slope $dV/dt$ at the 7 $\sigma_V$ threshold and the signal-to-noise ratio. 
All quantities were measured at ambient temperature of -35$^\circ$C, and at $P_{\it density}$ = 0.9 W/cm$^2$ for each analog pixel available and then averaged between the pixels that were subject to the same fluence (see Table \ref{tab:fluences}). 
The data displayed as black squares were collected at  HV = 200 V and $V_{\it CCA}$ = 1.8 V.
The green lines represent the average values measured with the 12 pixels of three unirradiated chips.

The measurements of Figure \ref{fig:fluencePlots} show that the signals are unaffected by radiation up to $6 \times 10^{14}$ n$_{\text{eq}}$/cm$^{2}$.
For larger fluence values, the average amplitude and $\sigma_V$ show a degradation, reducing the average signal-to-noise ratio up to a factor of three.
The signal slope seems to be somewhat less affected by radiation than the other signal characteristics.

For one of the two chips irradiated at $1 \times 10^{16}$ n$_{\text{eq}}$/cm$^{2}$, with a V$_{\it CCA}$ value supplied of 1.8 V, the signal-to-noise ratio of two of the four analog pixels was not large enough for a smooth operation.
This variation can probably be attributed to process mismatch. 
These two channels were recovered by increasing the $V_{\it CCA}$ of the chip to 2.0 V and the sensor bias to 250 V, which increased the signal amplitude and led to a large enough signal-to-noise ratio to operate them well. 
For the two highest fluence values, data were thus taken also at HV = 250 V and $V_{\it CCA}$ = 2.0 V (red dots). 
The improved performance at high proton fluence of this second working point shows that
$V_{\it CCA}$ and  the sensor bias represent two other parameters, in addition to $I_{\it fbk}$, that allow smooth and effective operation of SiGe HBT  frontend electronics even at a proton fluence of $1 \times 10^{16}$ n$_{\text{eq}}$/cm$^{2}$.

\section{Estimation of the time jitter of the signal}

The characterisation of a timing detector using a radioactive source has quite some differences and challenges with respect to a testbeam measurement. Indeed:
{\it i)} there is no control on the track selection that an external beam telescope would provide, with the consequence that, e.g., tracks at all angles with respect to the sensor are included in the sample analysed;
{\it ii)} there is no external time reference provided by fast detectors like, e.g., MPCs.
Although measurements with a radioactive source  cannot substitute testbeam measurements, they provide flexibility and can  bring precious information at the early stages of qualification of a timing detector.
For these reasons, data taken with a $^{90}$Sr source were used for a first assessment of the radiation tolerance of the prototype in SiGe BiCMOS technology presented in~\cite{Zambito_2023}. The time jitter of the signal was evaluated using the method described below, which utilises the time jitter computed at the most probable value (mode) of the signal amplitude distribution. The accuracy of this method for non irradiated chips was established 
using it to reanalise the testbeam data published in~\cite{Zambito_2023}
and  
by comparing the resulting time jitter with the time resolution that was obtained by time-of-flight method.

\subsection{Measurement method}
\label{subsec:jittermeasurement}

\begin{figure}[!htb]
    \centering
    \includegraphics[width=0.70\textwidth]{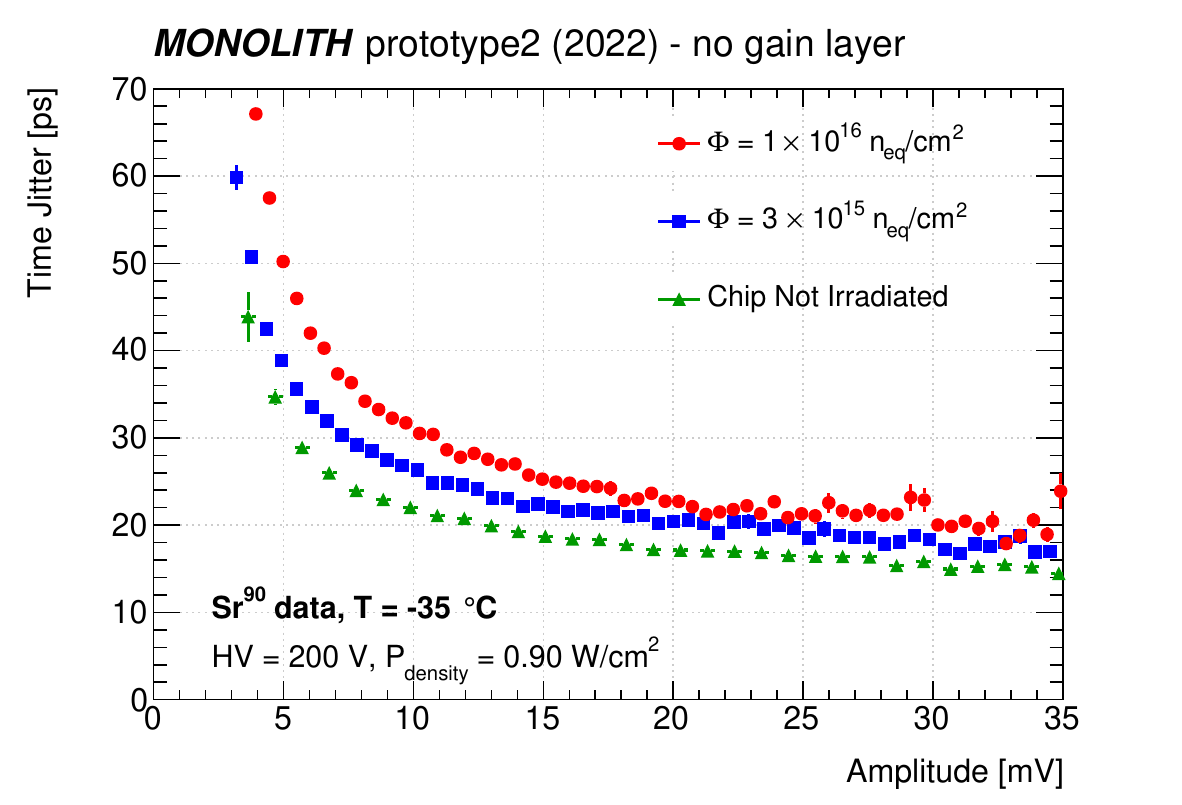}
    \caption{Average value of the time jitter calculated according to equation~\ref{eq:jitter} in each bin of the signal amplitude, measured with a $^{90}$Sr radioactive source for three different fluences. The data refer to one of the four analog pixels in each chip. 
    For the sake of clarity, the results obtained with the four chips irradiated up to $6 \times 10^{14}$ n$_{\text{eq}}$/cm$^{2}$ are not shown in the figure since they overlap to the distributions obtained for the unirradiated chip.
    }
    \label{fig:jitterAmplitude}
\end{figure}

For each signal, the time jitter measured with the $^{90}$Sr radioactive source was obtained as:
\begin{equation}
     \text{signal time jitter}  = \frac{\sigma_{V}}{dV/dt},
    \label{eq:jitter}
\end{equation}
where the signal slope $dV/dt$ was calculated by linear interpolation of the oscilloscope signal samplings between 6 $\sigma_V$ and the threshold of 7 $\sigma_{V}$ at which the signal arrival time was taken.
The time jitter of the most-probable signal amplitude, $\sigma_{t}^{^{90}\!{\text{Sr}}}$, was computed for each pixel in the following way:
\begin{enumerate}
    \item the values of the time jitter of all signals in the pixel, computed using formula~\ref{eq:jitter}, were plotted in bins of the signal amplitude, and the average values of the time jitter and the amplitude in each bin were retained.
    As example, Figure \ref{fig:jitterAmplitude} shows the result for a pixel of a chip not irradiated, one irradiated at $3 \times 10^{15}$  and one at  $1 \times 10^{16}$ n$_{\text{eq}}$/cm$^{2}$.
    \item then, the signal amplitude distribution was fitted with a Landau functional form, as shown in the bottom panels of Figure~\ref{fig:waveforms}, to obtain the mode of the Landau distribution; 
    \item finally, the value of the time jitter for the most-probable value $\sigma_{t}^{^{90}\!{\text{Sr}}}$ for the pixel was obtained by linear interpolation of the data points of the corresponding time jitter vs. amplitude plot, computed at the mode value of its amplitude distribution.
\end{enumerate}

The oscilloscope contribution to $\sigma_V$, which was typically 150 $\mu$V, was subtracted in quadrature.

To verify its level of accuracy, this method was also used to reanalyze the testbeam data reported in~\cite{Zambito_2023}. It was found that it provides a signal time jitter of (22.7 $\pm$ 0.3) ps to be compared with the time resolution of (23.8 $\pm$ 0.3) ps published  in~\cite{Zambito_2023}.
Therefore, we can conclude that the results obtained with the method utilised here to estimate the time resolution using the electrons from a $^{90}$Sr source are accurate at the 10\% level for the unirradiated boards.

\subsection{Time jitter vs. fluence}
\label{subsec:jitterVSfluence}



The time jitter vs. signal amplitude distributions of the three chips in Figure \ref{fig:jitterAmplitude} show the same trend, but the time jitters are systematically worse for increasing proton fluence, giving an indication of the level of degradation of the time resolution with exposition to increasing fluence. 
Plots analogous to those in Figure \ref{fig:jitterAmplitude} were produced for each of the 40 analog pixels of the ten chips used for this study, and were utilised to estimate the time jitter with the method described above.
%
\begin{table}[hbt]
\setlength{\tabcolsep}{10pt} 
\renewcommand{\arraystretch}{1.1}
\centering
\begin{tabular}{|c|cccc|c|}
\hline
\multirow{2}{*}{Fluence {[}n$_{\text{eq}}$/cm$^{2}${]}} & \multicolumn{4}{c|}{$\sigma_{t}^{^{90}\!{\text{Sr}}}$ {[}ps{]}} & \multirow{2}{*}{Average $\sigma_{t}^{^{90}\!{\text{Sr}}}$ {[}ps{]}} \\ & \multicolumn{1}{l}{pixel 1} & \multicolumn{1}{l}{pixel 2} & \multicolumn{1}{l}{pixel 3} & \multicolumn{1}{l|}{pixel 4} & \\
\hline
\multicolumn{6}{|c|}{HV = 200 V, $V_{\it CCA}$ = 1.8 V} \\ 
\hline
\multirow{3}{*}{0} & 22.1 & 20.5 & 18.8 & 19.9 & \multirow{3}{*}{21.0 $\pm$ 1.4} \\ 
& 22.1 & 22.7 & 19.5 & 19.6 & \\
& 22.9 & 22.2 & 21.1 & 20.7 & \\ 
\hline
$2 \times 10^{13}$ & 21.4 & 22.2 & 21.2 & 22.4 & 21.8 $\pm$ 0.6 \\
\hline
$9 \times 10^{13}$ & 21.4 & 22.5 & 21.0 & 21.8 & 21.7 $\pm$ 0.6 \\
\hline
\multirow{2}{*}{$6 \times 10^{14}$} & 21.5 & 22.4 & 20.2 & 20.9 & \multirow{2}{*}{21.5 $\pm$ 0.8} \\ 
& 20.7 & 22.3 & 22.6 & 21.3 & \\ 
\hline
$3 \times 10^{15}$ & 32.7 & 33.2 & 31.4 & 32.8 & 32.5 $\pm$ 0.8 \\
\hline
\multirow{2}{*}{$1 \times 10^{16}$} & 43.3 & 50.9 & 44.0 & 47.5 & \multirow{2}{*}{56.6 $\pm$ 16.3} \\
& 84.9 & 79.6 & 48.9 & 53.7 & \\ 
\hline
\multicolumn{6}{|c|}{HV = 250 V, $V_{\it CCA}$ = 2.0 V} \\ 
\hline
$3 \times 10^{15}$ & 28.7 & 29.0 & 28.5 & 29.5 & 28.9 $\pm$ 0.4 \\
\hline
\multirow{2}{*}{$1 \times 10^{16}$} & 33.2 & 36.2 & 35.5 & 33.6 & \multirow{2}{*}{39.6 $\pm$ 6.6} \\
& 51.4 & 46.9 & 38.4 & 41.7 & \\ 
\hline
\end{tabular}
\caption{Time jitter of the 40 analog pixels used for this study. A row reports the proton fluence at which the chip was exposed and the time jitter $\sigma_{t}^{^{90}\!{\text{Sr}}}$ measured for the 4 analog pixels in the chip.  The last column shows the time jitter obtained by averaging the pixels exposed to the same proton fluence; the error shown is the standard deviation of the values.
The data were collected using the single-ended signals produced by the electrons emitted by a $^{90}$Sr source. 
The last three rows report the time-jitter values obtained at HV = 250 V and $V_{\it CCA}$ = 2.0 V, which provide a better and more uniform time jitter for the various pixels.}
\label{tab:results} 
\end{table}

The top part of table~\ref{tab:results} reports the resulting time jitter of the 40  pixels, measured at the same working point.
The last column shows the average time jitter obtained for all pixels exposed to the same proton fluence.
The error quoted represents the standard deviation of the pixels used for the average.
The time jitter of the three unirradiated chips is $\sigma_{t}^{^{90}\!{\text{Sr}}}$ = (21.0 $\pm$  1.4) ps. This result is compatible within  errors with the time jitter obtained with the chips irradiated up to $6 \times 10^{14}$ n$_{\text{eq}}$/cm$^{2}$. 
For larger proton fluence, the time jitter at the same sensor bias voltage of HV = 200 V and $V_{\it CCA}$ = 1.8 V becomes $\sigma_{t}^{^{90}\!{\text{Sr}}}$ = (32.5 $\pm$ 0.8) ps at $3 \times 10^{15}$ n$_{\text{eq}}$/cm$^{2}$ and $\sigma_{t}^{^{90}\!{\text{Sr}}}$ = (56.6 $\pm$ 16.3) ps at $1 \times 10^{16}$ n$_{\text{eq}}$/cm$^{2}$. 

\begin{figure}[!htb]
    \centering
    \begin{minipage}[b]{0.70\textwidth}
        \includegraphics[width=\textwidth]{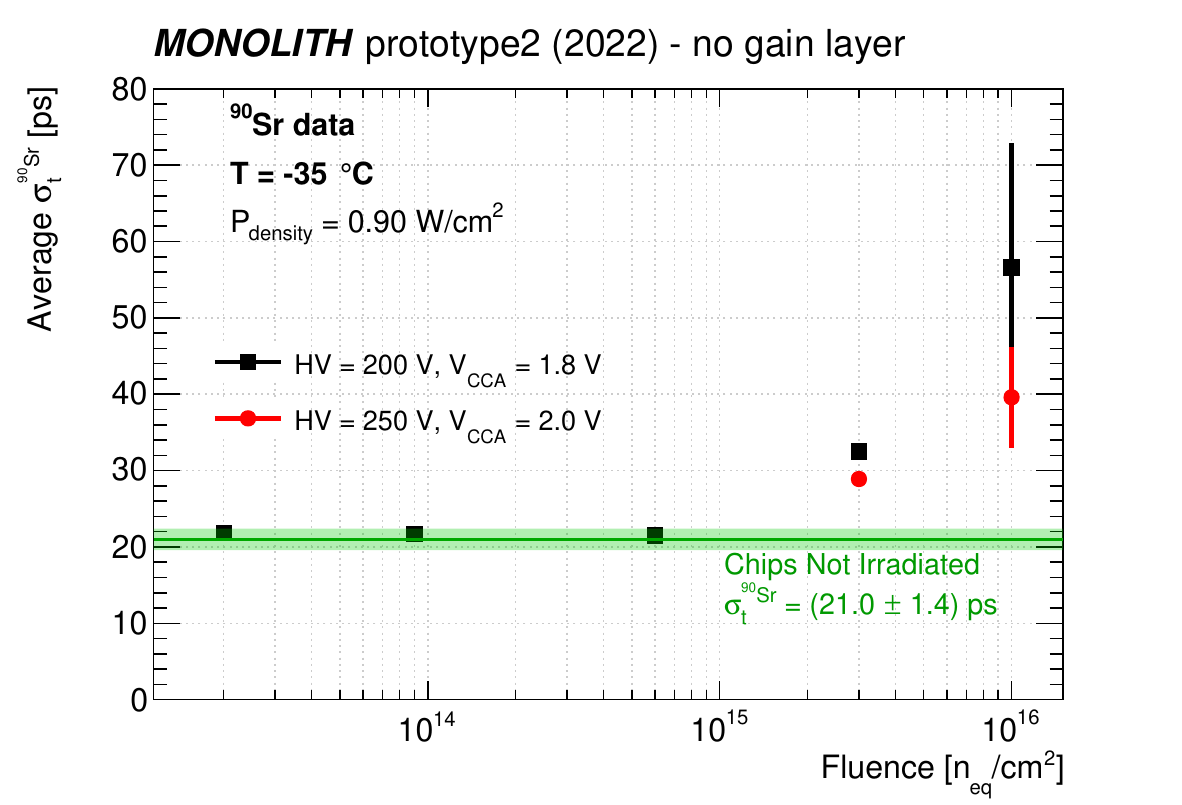}
    \end{minipage}
    \caption{Average time jitter $\sigma_{t}^{^{90}\!{\text{Sr}}}$ measured with the $^{90}$Sr radioactive source as a function of the fluence. 
    Data were collected at -35$^\circ$C and $P_{\it density}$ = 0.9 W/cm$^2$, and with two different working points: HV = 200 V and $V_{\it CCA}$ = 1.8 V (black squares), or HV = 250 V and $V_{\it CCA}$ = 2.0 V (red dots). 
    For each proton fluence, the time jitter was calculated independently for each analog pixel and then averaged. The error bars displayed represent the standard deviations of the average values.
    The green line represents the average value measured with the 12 analog pixels of the three unirradiated chips operated in the same conditions as the black squares; the green band represents the standard deviation of the average value.}
    \label{fig:TimeJitterPlots}
\end{figure}

Table~\ref{tab:results} shows that two pixels of one of the chips irradiated to $1 \times 10^{16}$ n$_{\text{eq}}$/cm$^{2}$ have time jitter of approximately 80 ps and are responsible for the large standard deviation at this fluence. As anticipated in section \ref{subsec:radioactiveSourceChoice},  to recover the performance of these channels the $V_{\it CCA}$ was increased to 2.0 V and the sensor bias to HV = 250 V. 
At this modified working point the average time jitters, reported in the bottom part of Table~\ref{tab:results}, become $\sigma_{t}^{^{90}\!{\text{Sr}}}$ = (28.9 $\pm$ 0.4) ps at $3 \times 10^{15}$ n$_{\text{eq}}$/cm$^{2}$ and $\sigma_{t}^{^{90}\!{\text{Sr}}}$ = (39.6 $\pm$ 6.6) ps at $1 \times 10^{16}$ n$_{\text{eq}}$/cm$^{2}$.
The average time jitters are also shown in Figure~\ref{fig:TimeJitterPlots} by black squares or red dots and for the unirradiated chips by the horizontal green bar.
Since the signal slope does not change significantly between different levels of fluence (as shown by the top-left panel of Figure \ref{fig:fluencePlots}), the worsening of the time jitter can be attributed mostly to the increase of the voltage noise $\sigma_V$ with increasing fluence reported in Figure \ref{fig:fluencePlots} bottom-left.



\subsection{Time jitter vs. sensor bias}

High level of radiation crossing a silicon sensor damages the substrate, generating a variation of the resistivity and an increase of the concentration of charge traps (\cite{Watkins_2000}, \cite{Bruzzi_2001}). These might lead to a visible change in the charge produced in the sensor and collected by the frontend after irradiation. Indeed:
\begin{itemize}
    \item if the resistivity of the substrate changes, the voltage needed to fully deplete a sensor before irradiation may not be sufficient to fully deplete the  sensor after irradiation, and the primary charge produced is less;
    \item the traps formed in the silicon substrate by radiation may capture part of the drifting charge so that the total charge collected by the preamplifier is less than the total charge that would be collected by a sensor not irradiated. 
\end{itemize}

Since both effects could be mitigated increasing the sensor bias voltage,
the consequence of irradiation on the sensor substrate was studied using data taken with the $^{90}$Sr source at different values of sensor bias voltage.
To single out
variations of performance 
that
could be attributed solely to the sensor and 
avoid
contributions from the electronics, the parameters of operation of the electronics were kept at the standard working-point values of 
$I_{\it preamp}$ = 50 $\mu$A,
$I_{\it fbk}$ = 2.0 $\mu$A
and
$V_{\it CCA}$ = 1.8 V
during the sensor bias voltage scan.  
\begin{figure}[!htb]
    \centering
    \includegraphics[width=0.70\textwidth]{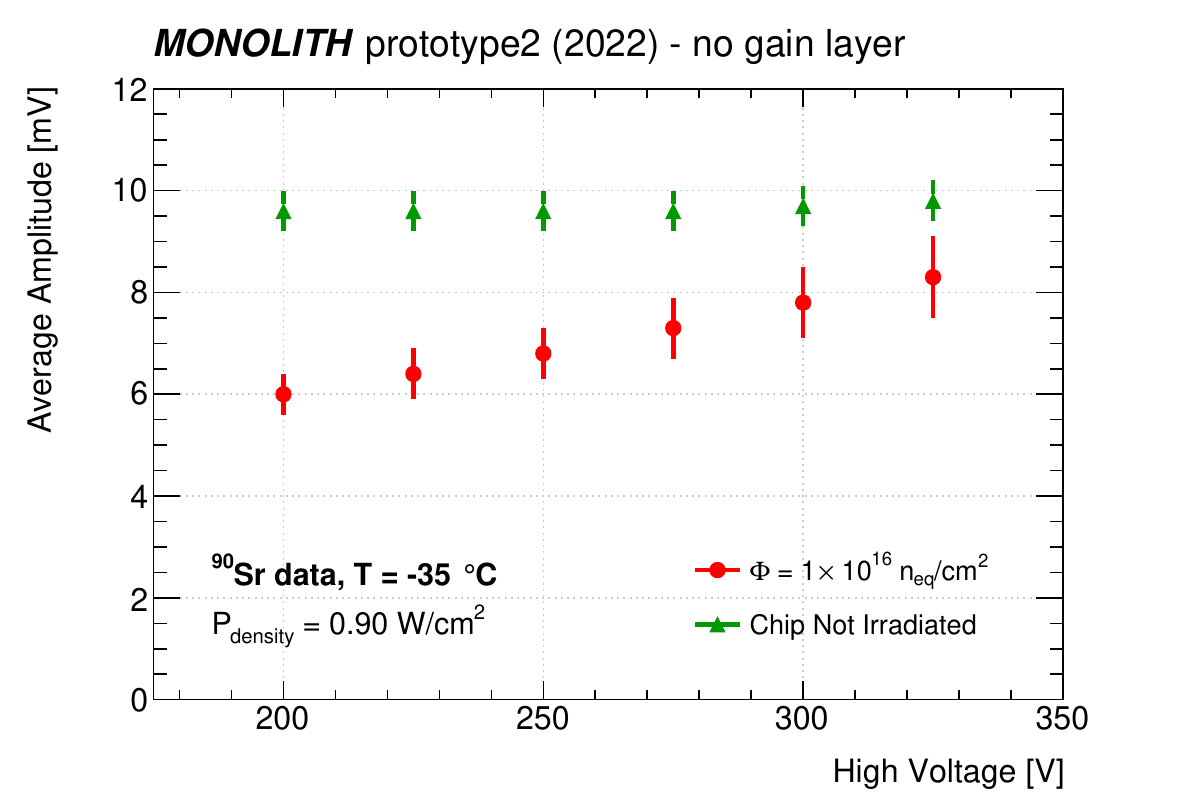}
    \caption{Average mode of the signal amplitude measured with $^{90}$Sr as a function of the sensor bias voltage for one of the three chips not irradiated and one of the two chips irradiated to $1 \times 10^{16}$ n$_{\text{eq}}$/cm$^{2}$. The values plotted represent the mode of the amplitude distributions obtained by fits using a Landau functional form. The amplitude was calculated independently for each of the four analog pixels in a chip and then averaged. The error bars displayed represent the standard deviation of the average values.
    }
    \label{fig:NormAmplitudes}
\end{figure}

The $^{90}$Sr source is an almost pure emitter of $\beta^-$ particles that deposit charge all along their path in the substrate; thus, the amount of charge produced is expected to increase for larger depleted volumes. 
In addition, an increase of the depleted volume with  increasing sensor bias voltage also decreases the capacity of the sensor,  leading to a higher signal amplitude for the same amount of charge. 
A concurrent reduction of  charge trapping 
with increasing bias voltage might also contribute to the measured trend.
%

Figure \ref{fig:NormAmplitudes} shows the dependence of the mode of the distribution of the signal amplitudes on the sensor bias voltage 
for the chip not irradiated of the first row of Table~\ref{tab:results} and  the chip irradiated to $1 \times 10^{16}$ n$_{\text{eq}}$/cm$^{2}$  of the ninth row of Table~\ref{tab:results}.
In the case of the chip not irradiated, the mode of the amplitude does not vary with the sensor bias voltage, which demonstrates that 200 V is enough bias to deplete the sensor fully.
On the other hand,
in the case of
the chip irradiated to $1 \times 10^{16}$ n$_{\text{eq}}$/cm$^{2}$ the  mode of the amplitude is found to increase with  increasing sensor bias; this observation suggests that increasing the sensor bias voltage above 200 V increases the depletion volume, thus producing more charge and reducing the capacitance, and reduces charge trapping.

\begin{figure}[!htb]
    \centering
    \includegraphics[width=0.70\textwidth]{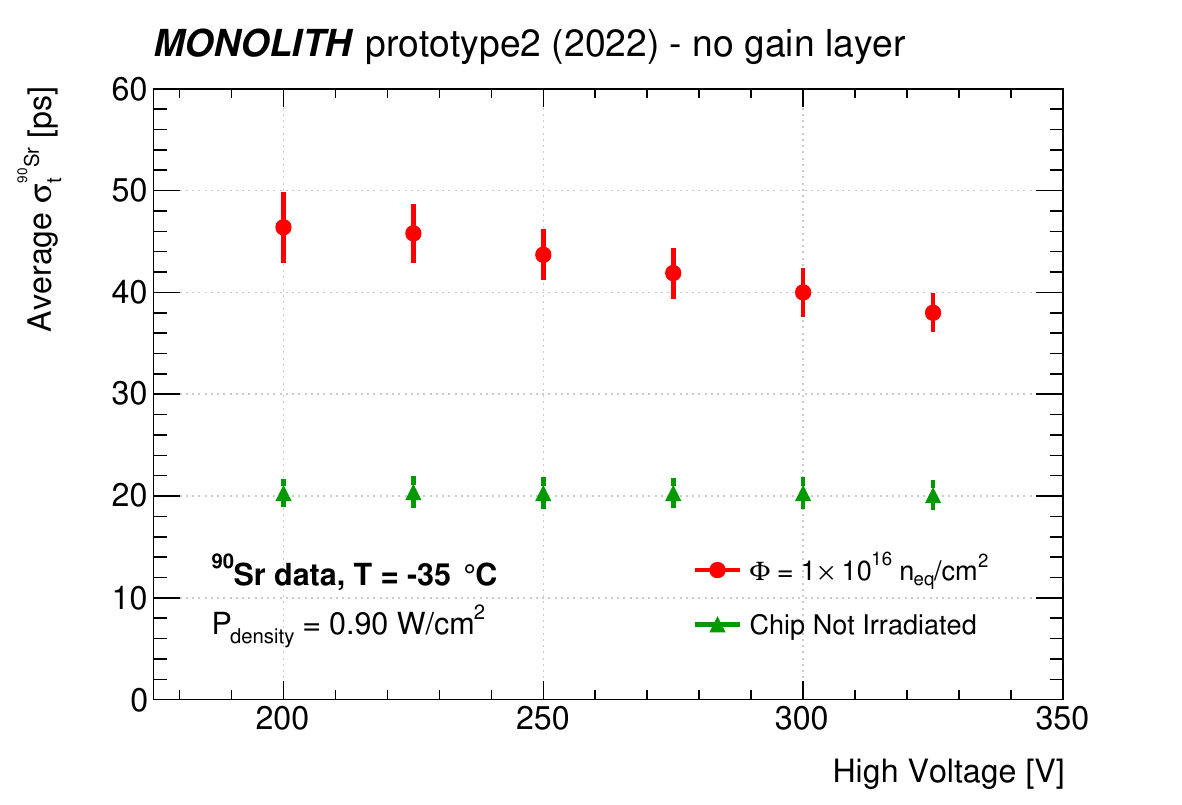}
    \caption{Average time jitter $\sigma_{t}^{^{90}\!{\text{Sr}}}$ measured with the $^{90}$Sr radioactive source as a function of the sensor bias voltage for one of the two chips irradiated to $1 \times 10^{16}$ n$_{\text{eq}}$/cm$^{2}$ (red dots) and for one of the unirradiated chips (green triangles). 
    Data were collected at -35$^\circ$C. The electronics was operated at the standard working point with 
    $I_{\it preamp}$ = 50 $\mu$A (corresponding to $P_{\it density}$ = 0.9 W/cm$^2$),
    $I_{\it fbk}$ = 2.0 $\mu$A
    and
    $V_{\it CCA}$ = 1.8 V. 
    The time jitter was calculated independently for each of the four analog pixels and then averaged. The error bars displayed represent the standard deviation of the average values.}
    \label{fig:HVScan}
\end{figure}

To quantify the effect of this change in signal amplitude on the timing performance, Figure \ref{fig:HVScan} shows the average time jitter of the most-probable signal amplitude $\sigma_{t}^{^{90}\!{\text{Sr}}}$ as a function of the sensor bias voltage, 
measured for the same two chips of Figure~\ref{fig:NormAmplitudes}.
When the sensor bias voltage is varied between 200 and 325 V, the value of $\sigma_{t}^{^{90}\!{\text{Sr}}}$  remains flat at approximately 20 ps for the chip not irradiated,
while  it gradually improves from  $\sigma_{t}^{^{90}\!{\text{Sr}}}$ = (46.4 $\pm$ 3.5) ps to (38.0 $\pm$ 1.9) ps for the chip irradiated to $1 \times 10^{16}$ n$_{\text{eq}}$/cm$^{2}$.

\section{Conclusions}
\label{sec:conclusions}

Samples of the monolithic SiGe BiCMOS ASIC prototype of the MONOLITH ERC Advanced project were irradiated at the CYRIC  facility in Japan with 70 MeV protons up to a maximum fluence of $1\times 10^{16}$ n$_{\text{eq}}$/cm$^{2}$.
After bypassing the radiation-damaged electronic components of the readout boards, the chips could be  operated.
Data were taken with a $^{90}$Sr radioactive source at a temperature of -35$^\circ$C, a sensor bias voltage of 200 V, and a power density of 0.9 W/cm$^2$, to characterize the irradiated boards together with three not irradiated boards used for reference.

Inspection of the pedestal-noise and  signal-amplitude distributions indicates that the favourable signal-to-noise ratio provided by SiGe BiCMOS technology permits to set the signal threshold low enough to operate the sensor at high efficiency even at a fluence of $1 \times 10^{16}$ n$_{\text{eq}}$/cm$^{2}$.

The time jitter of each signal was calculated as the ratio between the voltage noise $\sigma_V$ and the signal slope measured at the 7 $\sigma_V$ threshold, and the timing performance  of each pixel was estimated by the time jitter of the most-probable signal amplitude.
The results obtained with this method for the boards not irradiated reproduced the timing performance at the level of 20 ps that was measured at the testbeam.
No degradation of the 
timing performance
was observed up to proton fluence of $6 \times 10^{14}$ n$_{\text{eq}}$/cm$^{2}$.
At $1 \times 10^{16}$ n$_{\text{eq}}$/cm$^{2}$ the time jitter of the most-probable signal amplitude was found to be 
$\sigma_{t}^{^{90}\!{\text{Sr}}}$ = (56.6 $\pm$ 16.3) ps.
A new working point specific to  this proton fluence, 
that includes larger sensor bias voltage and larger analog voltage supplied to the preamplifier,
brings the time jitter to 
$\sigma_{t}^{^{90}\!{\text{Sr}}}$ = (39.6 $\pm$ 6.6) ps.

These results show that
SiGe BiCMOS processes can be considered for the production of very high time resolution pixelated silicon detectors without  internal gain layer for future colliders and other disciplines  involving very high radiation environments.

\acknowledgments
This research is supported by the Horizon 2020 MONOLITH  ERC Advanced Grant ID: 884447. The authors wish to thank Coralie Husi, Javier Mesa, Gabriel Pelleriti, and the technical staff of the University of Geneva and IHP Microelectronics.
The authors acknowledge the support of EUROPRACTICE in providing design tools and MPW fabrication services.
\newpage
\bibliographystyle{unsrt}
\bibliography{bibliography.bib}
\end{document}